\documentclass[aps,letterpaper,nofootinbib,preprint,showpacs,preprintnumbers,amsmath,amssymb]{revtex4}%
\usepackage{latexsym}%
\usepackage{hyperref}%

\def\cF{{\mathcal F}}
\def\cH{{\mathcal H}}

\def\cN{{\mathcal N}}

\def\cW{{\mathcal W}}

\newcommand{\beq}{\begin{equation}}
\newcommand{\beqn}{\begin{equation}\nonumber}
\newcommand{\eeq}{\end{equation}}
\newcommand{\bea}{\begin{eqnarray}}
\newcommand{\bean}{\begin{eqnarray}\nonumber}
\newcommand{\eea}{\end{eqnarray}}

\begin{document}

\begin{center}
{\bf {\Large Mass Spectrum and Statistical Entropy}}\\
\smallskip
{\bf {\Large of the BTZ black hole}}\\
\smallskip
{\bf {\Large from Canonical Quantum Gravity}}

\bigskip
\bigskip

{{Cenalo Vaz$^{a}$\footnote{e-mail address: Cenalo.Vaz@UC.Edu},
Sashideep Gutti$^{b}$\footnote{e-mail address: sashideep@tifr.res.in},
Claus Kiefer$^{c}$\footnote{e-mail address: kiefer@thp.uni-koeln.de},
T. P. Singh$^{b}$\footnote{e-mail address: tpsingh@tifr.res.in}, and\\
L.C.R. Wijewardhana$^{a}$\footnote{e-mail address: Rohana.Wijewardhana@UC.Edu}}}
\bigskip
\bigskip
\medskip

{\it $^{a}$RWC and Department of Physics, University of Cincinnati}\\
{\it Cincinnati, Ohio 45221-0011, USA}
\medskip

{\it $^{b}$Tata Institute of Fundamental Research,}\\
{\it Homi Bhabha Road, Mumbai 400 005, India}
\medskip

{\it $^{c}$Institut f\"ur Theoretische Physik, Universit\"at zu K\"oln}\\
{\it Z\"ulpicher Stra\ss e 77, 50937 K\"oln, Germany}
\end{center}
\medskip

\centerline{ABSTRACT}
\noindent
In a recent publication we developed a canonical quantization program describing the 
gravitational collapse of a spherical dust cloud in 2+1 dimensions with a negative cosmological 
constant $-\Lambda\equiv -l^{-2}<0$.  In this paper we address the quantization of the 
Banados--Teitelboim--Zanelli (BTZ) black hole. We show that the mass function describing the 
black hole is made of two pieces, a constant non-vanishing boundary contribution and a discrete 
spectrum of the form $\mu_n = \frac{\hbar}{l}\left(n+ \frac 12 \right)$. The discrete spectrum 
is obtained by applying the Wheeler--DeWitt equation with a particular choice of factor ordering 
and interpreted as giving the energy levels of the collapsed matter shells that form the black 
hole. Treating a black hole microstate as a particular distribution of shells among the levels, 
we determine the canonical entropy of the BTZ black hole. Comparison with the Bekenstein--Hawking 
entropy shows that the boundary energy is related to the central charge of the Virasoro algebra 
that generates the asymptotic symmetry group of the three-dimensional anti-de Sitter space AdS$_3$. 
This gives a connection between the Wheeler--DeWitt approach and the conformal field theory approach. 

\medskip

\vfill\eject
\section{Introduction\label{intro}}

The four laws of black hole mechanics formalized by Bardeen, Carter, and Hawking \cite{bch73} 
capture the macrosocopic properties of black holes and provide compelling evidence that black 
holes are thermodynamic systems. Each of the four laws has an analogue in the laws of 
thermodynamics if one identifies the surface gravity, $\kappa$, with some multiple of the 
temperature and the surface area of the horizon with the thermodynamic entropy. 

Even before the formulation of the four laws, Bekenstein \cite{bek72,bek72a} had proposed that 
black holes should possess an entropy given by some multiple of the horizon area since, 
according to the classical theory, there is no process by which the area of the horizon can 
decrease \cite{haw71}. Bekenstein's idea was later confirmed in an independent way by Hawking 
\cite{haw75} who showed that an eternal black hole behaves as a heat reservoir whose temperature 
is determined precisely by its surface gravity according to $T=\kappa/2\pi$. Hawking's result 
is consistent with Bekenstein's proposal if the multiple is taken to be $1/4$. It leads to the 
now famous Bekenstein-Hawking formula for the entropy of the black hole:
\beq
S_\text{B-H}=\frac{A}{4G\hbar}.
\label{arealaw}  
\eeq
For a solar mass black hole, this is on the order of $10^{77} k_B$, where 
$k_B$ is Boltzmann's constant, whereas the entropy of our sun is estimated to be about 
twenty orders of magnitude smaller. Thus, in collapsing to form a black hole, a matter cloud 
of one solar mass would gain entropy by a factor of $10^{20}$. Quantum gravity is expected 
to play a pivotal role in this increase of entropy \cite{OUP}.  

Because a successful theory of quantum gravity should be able to determine the microstates 
of a black hole, it had long been hoped that obtaining the area law \eqref{arealaw}
by counting the states in the sense of Boltzmann would serve as a powerful test of a candidate
theory and perhaps even to discriminate between different approaches. On the contrary, many 
apparently disparate models, from loop quantum gravity \cite{rov96,abck97,dle04,mei04,cdf06} 
to string theory \cite{stva96,hrpl97,dab05} and the AdS/CFT correspondence 
\cite{hkl99,emp99,hmst01,mupa02,chgs07}, have claimed partial success
in recovering \eqref{arealaw}. One could call this state of affairs an
``embarrassment of riches'' \cite{CarlipGRG}. 
The states being counted differ in each approach and the true underlying quantum gravitational 
states remain shrouded in mystery. Moreover, often the actual physical degrees of freedom being 
counted are themselves poorly understood and there are differences in the details of the counting 
of states, in particular in the statistical principles that are used. For instance, in loop 
quantum gravity the black hole microstates are represented by punctures of a spin network on the 
horizon. The counting of these states yields the area law \eqref{arealaw} provided that Boltzmann 
statistics (where the `particles' are assumed distinguishable) and not
Bose statistics (where the `particles' are assumed indistinguishable)
are implemented. On the other hand, in string theory the  
microstates of BPS black holes are dual to weak field D-brane states,
which are counted using Bose statistics. This difference in the
statistics is important for an understanding of the microscopic
degrees of freedom and for the calculation of logarithmic corrections
to \eqref{arealaw}, see \cite{Kolland}. 

Ever since its discovery, the Ba\~nados--Teitelboim--Zanelli (BTZ)
black hole \cite{btz92,bhtz93} has 
been an object of intense  
study both as a toy model for black-hole thermodynamics and as an aid to understanding the 
thermodynamics of many higher-dimensional black holes of interest to string theory, which 
can often be understood in terms of the BTZ solution. Soon after its discovery, Carlip 
suggested \cite{car95} that the BTZ black hole entropy could be understood in terms of the 
degrees of freedom of a Wess--Zumino--Witten (WZW) theory on the horizon. Not long after, 
Strominger \cite{str98} proposed 
that because the asymptotic symmetry group of 2+1-dimensional gravity with a negative 
cosmological constant is generated by two copies of the Virasoro algebra \cite{brhe86}, its 
degrees of freedom could be described by two conformal field theories
(CFTs) at infinity with central charges 
\beq
c_R = c_L = \frac{3l}{2G\hbar}.
\eeq
Most approaches to date apply CFT techniques to compute the BTZ black hole entropy. Yet,
more than a decade later the two most important open questions continue to be (a) what 
precisely are the degrees of freedom being counted and how are they related to the gravity 
theory that is being described, and (b) where do these degrees of
freedom ``lie'' (if they can be localized at all), that is,
whether they are asymptotic or whether they are excitations near the
horizon. Recently, Carlip  
has proposed a way to unify the two ideas by imposing the existence of a horizon as 
an additional constraint on the initial values of the gravitational degrees of freedom 
\cite{CarlipGRG,car05,car07}. The Hamiltonian and diffeomorphism constraints get modified by the 
``horizon constraint'' so that they generate a new Poisson bracket algebra that gives rise 
to the desired central charge.

Although canonical quantization of Einstein's general relativity cannot be expected to yield the 
final theory of quantum gravity, it is well adapted to address particular models with high
symmetry \cite{OUP}. It also has the virtue of ascribing a transparent
meaning to the canonical 
variables and therefore to the degrees of freedom. In canonical quantum gravity, the entropy of
a Schwarzschild black hole was understood in concrete terms as
encoding the unavailability, in the final 
collapsed state, of the information contained in the original matter distribution \cite{v00,vw01}. In this
paper we address the BTZ black hole in 2+1 dimensions from a similar point of view. Using a
program recently developed by us \cite{vgks07}, we now determine its microstates. Under certain natural
assumptions regarding factor ordering, we find that the energy spectrum coincides with that proposed 
in \cite{str98}. However, from the point of view of canonical gravity the meaning of the spectrum 
is clear: it represents the energy levels available
to shells of matter that have undergone collapse and now reside {\it within} the horizon. The mass 
energy of the BTZ black hole is the sum of the contribution from all the collapsed shells and a 
boundary term.\footnote{In 3+1-dimensional collapse, the boundary contribution 
from the origin is generally set to zero because a non-vanishing mass function at the origin would 
represent a singular initial configuration corresponding to a point mass at the center. The 
situation is quite different in the 2+1-dimensional models we are considering. A non-vanishing 
contribution from the origin is essential to allow for an initial velocity profile that vanishes 
there. This does not lead to singular initial data and the boundary contribution does not have 
the interpretation of a point mass situated at the center.} Explicitly counting the microstates 
yields an area law. The boundary term, which is arbitrary within the canonical theory, scales 
the entropy in much the same way as the central charge in the AdS/CFT approach. Its value can 
be determined by comparison with the Bekenstein-Hawking entropy in \eqref{arealaw}. Thus one 
gets a physical picture of the states being counted and also determines the connection between 
the boundary energy and the central charge of the Virasoro algebra
that generates the asymptotic AdS$_3$ symmetry.  

This paper is organized as follows. In Section II we briefly describe the models and their 
quantization, focusing only upon those aspects that are immediately related to our present purpose. 
For details on the classical solutions we refer the reader to \cite{roma,sg05}, for a semi-classical 
study of the Hawking radiation to \cite{tpg07}, and for more on the canonical quantization of 
these models we suggest \cite{vgks07}, from where most of Section II is taken. In Section III 
we focus on the BTZ black hole, which is a special case of the models described in section I. 
Here, the stationary bound states describing the BTZ black hole and its mass spectrum are obtained. 
In Section IV we obtain the entropy and conclude with some brief comments in Section V.

\section{The Models}

In \cite{vgks07} we considered the collapse of a spherical, inhomogeneous dust cloud in 2+1 
dimensions with a negative cosmological constant
$-\Lambda\equiv-l^{-2}<0$. These models are represented by a solution 
of Einstein's equations with pressureless dust described by the stress tensor $T_{\mu\nu} = 
\varepsilon U_\mu U_\nu$, where $\varepsilon(t,r)$ is the dust energy density. The solution is 
characterized by two arbitrary functions, 
the mass function $F(\rho)\equiv 4GM(\rho)$, representing the initial  
mass distribution and the energy function $E(\rho)$, representing the initial energy distribution
within the cloud. In terms of these functions, the classical solution is given as
\beq
ds^2 = -d\tau^2 + \frac{(\partial_\rho R)^2}{2(E-F)} d\rho^2 + R^2 d\varphi^2,
\label{classol}
\eeq
where $\tau$ is the dust proper time and $\rho$ labels shells of curvature radius $R(\tau,\rho)$.
The energy density $\varepsilon(\tau,\rho)$ is given by
\beq
8\pi G\varepsilon(\tau,\rho) = \frac{\partial_\rho F}{R (\partial_\rho R)},
\eeq
and Einstein's equations lead to 
\beq
(\partial_\tau R)^2 = 2E -\Lambda R^2,
\eeq
with solution
\beq
R(\tau,\rho) = \sqrt{\frac{2E}\Lambda}\sin\left(-\sqrt{\Lambda} \tau +\sin^{-1}\sqrt{
\frac{\Lambda}{2E}}~ \rho\right).
\label{Rsol}
\eeq
Shells labeled by $\rho$ become singular when $R(\tau,\rho)=0$. 

There is the freedom to choose the physical radius of a shell at the initial time. If 
$R(0,\rho) =\rho$, the mass and energy functions may be given in terms of the initial 
energy density of the collapsing cloud according to
\bea
F(\rho) &=& 8\pi G\int_0^\rho \rho' \varepsilon(0,\rho') d\rho' +f_0,\cr\cr
E(\rho) &=& [\partial_\tau R(0,\rho)]^2 + \Lambda \rho^2,
\eea
where $f_0$ is an arbitrary constant of integration. As $\partial_\tau R(0,\rho)$ is the 
initial velocity distribution, the first term in the expression for
$E(\rho)$ represents twice
the initial kinetic energy of the cloud.

The general circularly symmetric ADM metric may be embedded in the spacetime described by 
\eqref{classol}. In \cite{vgks07} we showed how, after a series of transformations, this leads to 
a canonical description in terms of a phase space consisting of $\tau(r)$, $R(r)$ and the 
mass density function, $\Gamma(r)$, defined via $F(r)$ according to
\beq
F(r) = \frac{M_0}2 + \int_0^r dr' \Gamma(r),
\label{massfn}
\eeq
where $M_0$ is an arbitrary constant, together with their canonical momenta, $P_\tau(r)$, 
$P_R(r)$ and $P_\Gamma(r)$, respectively; here and in the rest of this
section we set $G=1/8$ for convenience. 
The boundary actions at $r=0$ and $r\rightarrow 
\infty$ can be absorbed into a single hypersurface action, and the
effective canonical constraints of the gravity dust system are given as
\bea
\cH_r &=& \tau' P_\tau + R' P_R - \Gamma P_\Gamma' \approx 0\cr\cr
\cH &=& P_\tau^2 + \cF P_R^2 -\frac{\Gamma^2}{\cF} \approx 0,
\label{newconst}
\eea
where $\cF = \Lambda R^2 -2F$ and the prime represents derivation with respect to the ADM 
label coordinate $r$. The quantity $\cF$ is vanishing at the horizon, passing from 
negative in the interior of the black hole to positive in the exterior. This means that 
the DeWitt supermetric changes sign across the horizon and it is worth reexamining  
how this comes about. 

In \cite{vgks07} we start with a Hamiltonian constraint in which the gravitational part 
is hyperbolic everywhere and the matter part is only linear in momentum. The original 
canonical variables are sufficiently differentiable functions everywhere. However, they 
are not geometrically transparent and we perform a Kucha\v r (canonical) transformation 
\cite{kuc94} in the gravitational sector to a set of new variables, which are the mass function, 
$F$, the radial coordinate, $R$, and their conjugate momenta. Given in terms of the new 
variables, the Hamiltonian of the gravitational sector continues hyperbolic and is valid 
everywhere except at the horizon where it is degenerate and where the new canonical momenta 
generically suffer a jump. Introducing the new variable $\Gamma$, as defined in \eqref{massfn}, 
we arrive at yet another form of the gravitational Hamiltonian in which the kinetic term 
is of non-canonical form (containing the derivative of the momentum conjugate to $\Gamma$) 
and the matter part is still linear in the momentum. The advantage of this procedure is 
that it allows us to absorb the boundary terms into the hypersurface action. We then use 
the momentum constraint and square the Hamiltonian constraint of the gravity matter 
system to arrive at the final form \eqref{newconst}, where the kinetic term is hyperbolic 
inside and elliptic outside the horizon. The new constraint is thus a consequence of 
canonical transformations, use of the momentum constraint, and squaring. This sequence of
transformations does not preserve the hyperbolic form of the (gravitational part of the) 
original constraints, but the final constraints nevertheless are valid everywhere except 
at the horizon. We thus obtain the effective constraint system \eqref{newconst} and use it 
as a starting point for our discussion. In the quantum theory, the horizon will be treated 
as a boundary at which continuity and differentiability of the wave-functional are required. 

Dirac's quantization procedure may now be applied to turn the classical constraints 
into operator constraints on wave-functionals. According to it, the momenta 
are replaced by functional differential operators (we set $\hbar=1$)
\beq
P_X = -i  \frac{\delta}{\delta X(r)},
\label{diracq}
\eeq
and one may write the quantum Hamiltonian constraint as \cite{kmhv05}
\beq
{\widehat \cH} \Psi[\tau,R,\Gamma] = \left[\frac{\delta^2}{\delta \tau^2} + 
\cF \frac{\delta^2}{\delta R^2} + A\delta(0) \frac{\delta}{\delta R} + B\delta(0)^2+
\frac{\Gamma^2}{\cF}\right] \Psi[\tau,R,\Gamma] = 0,
\label{qham}
\eeq
where $A(R,F)$ and $B(R,F)$ are smooth functions of $R$ and $F$ which encapsulate the factor 
ordering ambiguities. The divergent quantities $\delta(0)$ and $\delta(0)^2$ are introduced 
to indicate that the factor ordering problem can be dealt with only after a suitable 
regularization procedure has been implemented. We notice that the Hamiltonian constraint 
contains no functional derivative with respect to the mass density function. In fact the mass 
density appears merely as a multiplier of the potential term in the Wheeler-DeWitt equation. 
This indicates that $\Gamma(r)$, and hence the initial energy density distribution, 
$\varepsilon(0,r)$ may be externally specified. Once specified, $\Gamma(r)$ determines the 
quantum theory of a particular classical model. 

The quantum momentum constraint, on the other hand, 
\beq
{\widehat \cH}_r \Psi[\tau,R,\Gamma] = \left[\tau'\frac{\delta}{\delta \tau} + R'\frac{\delta}
{\delta R} - \Gamma \left(\frac{\delta}{\delta \Gamma}\right)' \right] \Psi[\tau,R,\Gamma] = 0,
\label{qdiff}
\eeq
requires no immediate regularization because it involves only first order functional derivatives.
To describe a collapsing cloud with a smooth, non-vanishing matter density distribution over 
some label set of non-zero measure the Hamiltonian constraint was regularized on a lattice. 
The continuum limit of the wave-functional was taken to be of the form
\beq
\Psi[\tau,R,\Gamma] = \exp\left[i\int dr \Gamma(r) \cW(\tau(r),R(r),F(r))\right].
\label{wfnal}
\eeq
It automatically obeys the momentum constraint provided that $\cW(\tau,R,F)$ has no explicit 
dependence on the label coordinate $r$. We showed in \cite{vgks07} and \cite{kmhv05} that, 
for the wave-functionals to be simultaneously factorizable on the lattice and to obey the 
momentum constraint in the continuum limit (as the lattice spacing is made to approach zero), they 
must satisfy not one but three equations, one of which is the Hamilton--Jacobi equation that was used 
in earlier studies \cite{vksw03} to describe Hawking radiation in the
WKB approximation. The function $B(R,F)$ in \eqref{qham} is forced 
to be identically vanishing and the remaining two equations together with hermiticity of the 
Hamiltonian constraint uniquely fixed $A(R,F)$, the measure and the wave-functionals. Hermiticity 
of the operator
\beq
\widehat{\cF P_R^2} = -\cF(R,F) \frac{\partial^2}{\partial R^2} - A(R,F) \frac{\partial}
{\partial R} - B(R,F)
\eeq
on the lattice requires that 
\beq
A = |\cF|\partial_{R} \ln (\mu|\cF|),
\label{meas}
\eeq
where $\mu(R,F)$ is the measure that defines an inner product on the Hilbert space. Lattice 
regularization effectively turns the continuum (midi-superspace) problem into a countably 
infinite set of decoupled mini-superspace problems; the three equations mentioned earlier are 
required to ensure a sensible, diffeomorphism invariant continuum limit. 

If the mass density function is distributional in character (here it
is non-vanishing on a label set of 
measure zero to begin with), the wave-functional \eqref{wfnal} is automatically a wave-function, 
or a countable product of wave-functions, and the functional differential equations become 
ordinary partial differential equations. The original midi-superspace problem then naturally 
collapses into a set of mini-superspace problems from the very
beginning, and no further conditions must be met. The 
factor ordering ambiguities survive, of course, so there are not enough constraints to 
fully resolve the coefficients $A(R,F)$ and $B(R,F)$ together with the Hilbert space measure.
This is the case of the BTZ black hole, which we discuss in the following section.

\section{Stationary states representing the BTZ black hole}

As mentioned in the concluding paragraph of the previous section, the description of 
the eternal black hole differs fundamentally from our description of collapse in \cite{vgks07} 
because neither the coefficients $A(R,F)$ and $B(R,F)$ nor the Hilbert space measure are 
determined uniquely.  They must be chosen by different physical arguments, which we present 
here.

The BTZ black hole with ADM mass parameter $M$ is a special case of the solution \eqref{classol}. 
Reintroducing Newton's constant, $G$, which will facilitate comparison with other work, it is 
recovered when the mass function is taken to be constant, $F=4GM$, for $\rho>0$, and the energy 
function is given by $2E=1+8GM$, again as long as $\rho>0$. The metric in \eqref{classol} may then 
be brought to canonical (static) form, 
\beq
ds^2 = - \left(\Lambda R^2-8GM\right) dT^2 + \frac{dR^2}{(\Lambda R^2 -8GM)} + R^2 d\varphi^2,
\label{btzmetric}
\eeq
by the transformations $R=R(\tau,\rho)$ as given in \eqref{Rsol} and
\beq
T = \tau + \int dR \frac{\sqrt{1+8GM-\Lambda R^2}}{\Lambda R^2-8GM}
\eeq
for the Killing time, $T$. 

Within the framework of the canonical theory described in the previous section, the BTZ black
hole should be described by a mass function [see \eqref{massfn}] of the general form
\beq
F(r) = 4GM_0+4G\mu \Theta(r) = 4GM,
\label{massfn2}
\eeq
where $\mu$ represents the mass of a shell at $r=0$ and where $\Theta$ is the Heaviside function. 
Likewise, the energy function should be given by
\beq
E(r) = \frac 12 \left[1+8GM_0+8G\mu\Theta(r)\right].
\label{efn2}
\eeq
The mass function in \eqref{massfn2} yields a mass density that is the
$\delta-$distribution (recall that $F'(r)=\Gamma(r)$)
\beq
\Gamma(r) = 4G\mu\delta(r),
\eeq
and the wave-functional in \eqref{wfnal} turns into the wave-{\it function}, 
\beq
\Psi = e^{\frac i{4G}\int_0^\infty dr \Gamma(r) \cW(\tau(r),R(r),F(r))} = e^{i\mu \cW(\tau,R,F)},
\eeq
where $\tau=\tau(0)$, $R=R(0)$ and $F=F(0)$. The Wheeler--DeWitt
equation becomes 
the Klein--Gordon equation describing a shell of mass $\mu$,
\beq
\left[\frac{\partial^2}{\partial \tau^2}+\cF\frac{\partial^2}{\partial R^2} + A
\frac{\partial}{\partial R}+B\right]e^{i\mu\cW(\tau,R,F)} = 0,
\label{wd1}
\eeq
where we have absorbed the term $16\mu^2/\cF$ into $B$, which here
renormalizes the potential. 

As there are no further conditions, the choice of $A(R,F)$ and $B(R,F)$ for the BTZ black hole 
will necessarily be to some extent {\it ad hoc}. Because  
we are describing a single shell in this simple quantum mechanical
model of the black hole, we demand that 
\eqref{wd1} be the free wave equation. This is, of course, an
assumption, although a natural one. 
The free wave equation would be obtained if we could write
\eqref{wd1} as
\beq
\gamma^{ab} \nabla_a \nabla_b \Psi = 0,
\label{kg1}
\eeq
where $\gamma_{ab}$ is the DeWitt supermetric on the effective configuration space $(\tau,R)$
and $\nabla_a$ represents the covariant derivative on this space with respect to $\gamma_{ab}$. 
The supermetric is non-degenerate and can be read off directly from \eqref{newconst}. It is found 
to be flat, while being positive definite when $\cF>0$ and indefinite
when $\cF<0$, 
\beq
\gamma_{ab} = \left(\begin{matrix}
1 & 0\cr
0 & \frac 1\cF
\end{matrix}\right).
\eeq
The Wheeler--DeWitt equation \eqref{wd1} is therefore the free Klein--Gordon equation if we take 
$B(R,F)=0$,
\beq
A(R,F) = |\cF|\partial_R\ln \sqrt{|\cF|}
\label{AA}
\eeq
and an inner product given by the integral
\beq
\langle \Psi_1,\Psi_2\rangle = \int \frac{dR}{\sqrt{|\cF|}}~ \Psi_1^*~ \Psi_2,
\eeq
since inserting \eqref{AA} into \eqref{meas} gives  $\mu(R,F)=1/\sqrt{|\cF|}$.

When $\cF\neq 0$, the supermetric can be brought to a manifestly flat form by the coordinate 
transformation 
\beq
R_* = \pm \int \frac{dR}{\sqrt{|\cF|}},
\label{r*}
\eeq
and in terms of $R_*$, \eqref{kg1} reads
\beq
\left[\frac{\partial^2}{\partial \tau^2}\pm\frac{\partial^2}{\partial R_*^2}\right]
e^{i\mu\cW(\tau,R_*,F)} = 0,
\label{theeq}
\eeq
where in both the above equations, the upper sign refers to the exterior, $\cF>0$, and the lower 
to the interior, $\cF<0$. We note that the equation is hyperbolic in the interior, but elliptic 
in the exterior. This signature change has been noted in other models \cite{bk97} and comes about because 
$\cF$ passes from positive outside the horizon to zero on the horizon and negative inside. For 
our particular problem, it corresponds to the fact that the degrees of
freedom live {\it inside} the horizon. 

For the interior we find
\beq
R_*^\text{in} = \frac 1{\sqrt{\Lambda}} \sin^{-1} \sqrt{\frac{\Lambda R^2}{8GM}},
\eeq
so that the value of $R_*^\text{in}$ on the horizon is $\pi/(2\sqrt{\Lambda})$. In 
the exterior 
\beq
R_*^\text{out} =  \frac 1{\sqrt{\Lambda}} \left[\ln\left\{\frac{R\sqrt{\Lambda}+
\sqrt{\Lambda R^2-8GM}}{\sqrt{8GM}}\right\} + \frac\pi 2\right],
\eeq
where we have adjusted the integration constant so that $R_*$ is continuous across 
the horizon. If we extend its range to $(-\infty, \infty)$ so as to avoid any 
issues connected with a boundary at $R_*=0$, the solutions to the wave equation
may be given as
\beq
\left\{\begin{matrix}
\psi^\text{in}(\tau,R_*) = A_\pm e^{-i\mu(\tau\pm R_*)} & \cF<0\cr\cr
\psi^\text{out}(\tau,R_*) = B_\pm e^{-i\mu(\tau\pm i R_*)} & \cF > 0.
\end{matrix}\right.
\eeq
The ``interior'' is now the interval $(-\pi/2\sqrt{\Lambda},+\pi/2\sqrt{\Lambda})$ and
our solutions are oscillatory only in the interior. For a continuous and differentiable 
wave-function, matching conditions across the horizon require that
\beq
\mu_j = \frac{\hbar}{l}\left(j+\frac 12\right),~~ j \in \{0\}\cup \mathbb{N},
\label{spec}
\eeq
where $l=1/\sqrt{\Lambda}$ is again the AdS length, and we have
disallowed all negative values of $\mu$. We also re-insert now $\hbar$ in the 
expressions. Equation \eqref{spec} gives the allowed masses of a collapsed dust shell. 

We remark that a similar spectrum was proposed in \cite{str98} using arguments related to the 
asymptotic symmetry group of AdS$_3$, which is generated by (two copies of) the 
Virasoro algebra. In our current framework the CFT states get the interpretation of 
excited mass shells. Worthy of note is the fact that their spectrum has here been 
derived from the dynamics of the {\it interior of the hole}.

\section{Entropy}
To understand how the degeneracy that leads to the black hole entropy comes about, we 
propose the following point of view. We think of the black hole as a single shell given
by the mass and energy functions of \eqref{massfn2} and \eqref{efn2} respectively, with 
the well defined spectrum in \eqref{spec}. However, we know that this single shell 
is in fact the end-state of the collapse of many shells. Suppose that $\cN$ such shells
were to collapse to form the black hole. We assume that, regardless of their history, 
each can eventually occupy only the energy levels of \eqref{spec} upon collapse into the 
single shell. Thus, the black hole entropy is just the number of possible distributions 
of $\cN$ identical objects between these levels. In this way we realize Bekenstein's 
original ideas in \cite{bek73}. To quote from this article: ``It is then natural to 
introduce the concept of black-hole entropy as the measure of the {\it inaccessibility} 
of information (to an exterior observer) as to which particular internal configuration 
of the black hole is actually realized in a given case.'' 

A black hole microstate is therefore viewed as a particular distribution of shells among 
the available energy levels. In a particular microstate, let $\cN_j$ shells occupy level 
$j$. Taking into account the boundary contribution, $M_0$, the black hole's total mass 
becomes expressed in terms of the distribution of shells as
\beq
M = M_0 + \frac{\hbar}{l}\sum_j\left(j+\frac 12\right)\cN_j,
\eeq
and the BTZ solution in \eqref{btzmetric} is interpreted as an excitation by collapsed 
shells about some ground state solution. In the following section, we count the number 
of possible microstates. By requiring agreement between the entropy derived from this 
point of view and the Bekenstein--Hawking entropy we will then relate $M_0$ to the energy 
of the ground state.

Consider the canonical ensemble described by the partition function
\beq
Z(\beta) = \sum_{\{\cN_1,\ldots,\cN_j,\ldots\}}g(\cN_1,\ldots,\cN_j,\ldots) \exp
\left[-\beta\left(M_0+ \sum_{j}\mu_j \cN_j\right)\right],
\eeq
where $\cN_j$ represents the number of shells excited to level $j$, with mass $\mu_j$, 
and $g(\cN_1,\ldots,\cN_j,\ldots)$ represents the degeneracy of states. Our strategy will be 
to compute the canonical entropy 
\beq
\left. S_\text{can} = [\beta M + \ln Z(\beta)\right]_{M=-\partial \ln Z/\partial \beta},
\label{ent1}
\eeq
where $M$ is the average energy in the canonical ensemble, which we
associate with the black-hole mass. Both the canonical and the
microcanonical ensembles are well defined for the BTZ black hole
\cite{Reznik}. 
We implement bosonic statistics, $g(\cN_1,\ldots,\cN_j,\ldots)=1$. It is not 
difficult to see that Boltzmann statistics leads to the wrong
dependence of the entropy on the mass, see Appendix B. Again, it is
crucial to perform the counting of states in an appropriate way
\cite{Kolland}.

Interchanging the product with the sum and performing the 
sum over shells, we obtain
\beq
Z(\beta\hbar/2l) = e^{-\beta
  M_0}\prod_{j=0}^\infty\left[1-e^{-\frac{\beta\hbar}{2l}
    \left(2j+1\right)} 
\right]^{-1}.
\eeq
Using the remarkable identity \cite{hr18}
\beq
Z_0(\xi) = \prod_{j=1}^\infty \left[1-e^{-\xi j}\right]^{-1} =
\sqrt{\frac\xi{2\pi}}  
e^{\frac{\pi^2}{6\xi}-\frac{\xi}{24}}Z_0(4\pi^2/\xi),
\eeq
which follows from the Poisson summation formula, we show in Appendix A that
\beq
Z(\beta\hbar/2l) = \frac 1{\sqrt{2}}e^{-\left(\frac{\pi^2
      l}{\beta\hbar}+\frac{\beta\hbar}{8l}\right)\left( 
\frac{8lM_0}{\hbar}-\frac 16\right)} [Z(4\pi^2 l/\beta\hbar)]^{-1},
\label{modular}
\eeq
thus connecting the high temperature behavior of our system to its low
temperature dynamics. Since the temperature of the BTZ black hole
becomes large in the semi-classical limit of large mass (this is
different from the four-dimensional case), 
we will be interested in the high-temperature limit, $\beta
\rightarrow 0$; we must, therefore, determine 
the partition function $Z(4\pi^2 l/\beta\hbar)$ in the low-temperature
limit, that is, in the limit of infinite argument.  
The result depends on what we take to be the lowest energy state. If we take it
to be the $M=0$ state, described by the metric
\beq
ds^2 = -\frac{R^2}{l^2} dT^2 + \frac{l^2}{R^2} dR^2 + R^2 d\varphi^2,
\eeq 
then we get $Z(4\pi^2 l/\beta\hbar)\rightarrow 1$ in the limit of
infinite argument. From \eqref{modular} we then find the desired
result for large temperature, 
\beq
\ln Z(\beta\hbar/2l) \approx \frac{\pi^2l}{\beta\hbar} 
\left(\frac 16-\frac{8lM_0}{\hbar}\right).
\eeq
Therefore, when the temperature is large, the average energy of the system is
\beq
M = -\frac{\partial \ln Z}{\partial\beta}= \frac{\pi^2l}{\beta^2\hbar} \left(\frac 16-8lM_0\right),
\label{entemp}
\eeq
and we must require that $M_0\leq \hbar/48l = \hbar\sqrt{\Lambda}/48$ for a
positive mass. When this condition on $M_0$ holds, inserting
\beq
\beta \approx \pi\sqrt{\frac{l}{\hbar M}\left(\frac 16 -
    \frac{8lM_0}{\hbar}\right)} 
\label{tempmass}
\eeq
into \eqref{ent1} yields the canonical entropy 
\beq
S_\text{can} \approx
2\pi\sqrt{\left(1-\frac{48lM_0}{\hbar}\right)\frac{lM}{6\hbar}}. 
\label{scan}
\eeq
Using $R_h=l\sqrt{8GM}$ for the horizon of the BTZ black hole, the
Bekenstein--Hawking entropy in \eqref{arealaw}, 
\beq
S_\text{B-H} = \frac{\pi R_h}{2G\hbar} = \frac{\pi l}{\hbar}\sqrt{\frac{2M}G},
\eeq
coincides with the canonical entropy provided that $M_0$ takes the value
\beq
M_0 = - \frac 1{16G} + \frac{\hbar}{48l}.
\label{M01}
\eeq
This is compatible with our requirement that $M_0\leq \hbar/48l$. In
the semi-classical regime  
the cosmological constant must be small in Planck units, $l\gg G$;
therefore, $M_0$ is negative but larger than $-1/16G$.  

More generally, one would assign the value $\Delta_0$ to the energy of
the ground state; then, as $\beta \rightarrow 0$,
\beq
Z(4\pi^2 l/\beta\hbar) \approx e^{-\frac{8\pi^2 l^2\Delta_0}{\beta\hbar^2}}
\eeq
and therefore, using \eqref{modular},
\beq
\ln Z(\beta/2l\hbar) \approx \frac{\pi^2 l}{\beta\hbar} 
\left[\frac 16 - \frac{8l}{\hbar}(M_0-\Delta_0)\right].
\eeq
All the previous formul\ae\ apply after shifting $M_0$ to
$M_\text{eff}=M_0-\Delta_0$. 
Thus the energy of the vacuum solution is determined by the value of
$M_0$ according to 
\beq
\Delta_0 = M_0 + \frac 1{16G} - \frac{\hbar}{48l}.
\label{Deltazero}
\eeq
For example,
\beq
M_0=-\frac{3}{16G}+\frac{\hbar}{48l}
\eeq
gives $\Delta_0=-1/8G$, which corresponds to the choice of pure AdS$_3$,
\beq
ds^2= -\left(\frac{R^2}{l^2}+1\right)dT^2 + \left(\frac{R^2}{l^2}+1\right)^{-1}
dR^2 + R^2d\varphi^2,
\label{M02}
\eeq
for a ground state \cite{bhtz93}. 
For $-1/8G < \Delta_0 < 0$, \eqref{btzmetric} describes naked 
singularities. Even so, except for the fact that they are unstable,
there seems to  
be no impediment to choosing such solutions for the ground state. 

\section{Discussion}

Our approach to defining and computing the number of microstates of the BTZ black hole 
has been significantly different from those taken so far. While most approaches use the 
AdS/CFT correspondence, we apply only standard canonical techniques and the canonical 
variables that were adapted to the problem of gravitational collapse in \cite{vgks07}. The 
advantage 
of our approach is that the microstates have a transparent physical meaning. As such, we 
can identify the states being counted and make a connection with approaches that use CFT. 
In this section we will summarize our results and make this connection more precise. 

The black hole is viewed as a stationary state made up of collapsed shells. We obtained 
the energy levels available to the shells and found that these are the same as the 
levels expected from an asymptotic CFT. These energy levels were obtained subject to
a physical requirement concerning the factor ordering ambiguities that plague the 
canonical approach, that is, we required the wave equation to be the free (Klein--Gordon)
equation on the configuration space determined by the Hamiltonian constraint. Because the 
matter density describing a black hole is distributional we know of no {\it a priori}
determination of the factor ordering. This means that other factor orderings are possible,
and therefore other energy spectra. We do not know how, or even if, different factor 
orderings would be connected. This problem does not arise for genuine collapse scenarios,
where the matter distribution is taken to be smooth over some interval of the ADM label
coordinate, $r$. There, the diffeomorphism constraint uniquely fixed the factor ordering 
as well as the measure on the Hilbert space of states. It is therefore possible that,
once the quantum collapse process is more fully understood, this uniqueness will carry 
over to a unique description of the end state. This seems to be a worthy direction for 
future research.

The actual microstates of the black hole consist of distinguishable distributions of matter 
shells amongst the energy levels we have determined. The question of where the states lie, 
whether near the horizon or at infinity, is a misleading one in quantum mechanics. The 
canonical theory only distinguishes between two regions, the interior and the exterior, and 
the microstates of our model are shown to live {\it in the interior} of the hole, where the 
Wheeler--DeWitt equation is hyperbolic. The same picture holds for the Schwarzschild black hole 
in 3+1 dimensions, as argued in \cite{vw01}. However, there is a key difference between the 
BTZ black hole and its Schwarzschild counterpart. The BTZ black hole has positive specific 
heat, whereas the Schwarzschild black hole has negative specific heat. Thus the temperature of 
the BTZ black hole increases with its mass and the semi-classical approximation is also the 
high temperature limit. The Schwarzschild black hole, on the other hand, has negative specific 
heat, implying that the semi-classical approximation is in fact the low temperature limit.

Bose statistics were essential to correctly counting the number of
these microstates, cf. \cite{Kolland}.  We exploited a well known property of a certain partition 
function to obtain the number of states in the high temperature (large mass) limit. Our methods 
are no different from those used in CFT \cite{card86,card86a}. The entropy of the black hole depends 
on two parameters, {\it viz.,} 
the energy, $\Delta_0$, of the vacuum solution and a constant, $M_0$, arising from the 
boundary action at the origin of coordinates. Our result in \eqref{scan} compares directly with the 
answer obtained via the CFT approach \cite{str98},
\beq
S_\text{CFT} = 4\pi \sqrt{c_\text{eff} \frac{lM}{6\hbar}}
\eeq
for the zero angular momentum case and allows us to identify the quantity
\beq
\frac 12 \left[1-\frac{48l}{\hbar}(M_0-\Delta_0)\right]
\eeq
with the central charge, $c_\text{eff}$, of the effective CFT being used to describe the hole. 
To achieve agreement with the Bekenstein--Hawking entropy, we must choose $M_0-\Delta_0$ 
according to \eqref{Deltazero}. This leads to
\beq
c_\text{eff} = \frac{3l}{2G\hbar},
\eeq
which is the central charge of the Liouville theory induced at spatial infinity by 2+1-dimensional gravity \cite{chedr95}. On the other hand, Boltzmann statistics were employed in
\cite{vw01} to obtain the statistical entropy of the Schwarzschild black hole. It is 
desirable to have a deeper understanding of the relationship between the statistics and the 
presence or not of a cosmological constant as well as the dimensionality of the space-time
\cite{tgsv07}. We leave this important issue for future research.
\bigskip

\noindent{\bf Acknowledgements}
\bigskip

\noindent TPS gratefully acknowledges support from the German Science Foundation 
(DFG) under the grant 446 IND 113/34/0-1. LCRW was supported in part by the 
U.S. Department of Energy Grant No. DE-FG02-84ER40153. 
\bigskip\bigskip

\centerline{\bf APPENDIX A}
\bigskip

Here we show the details of the calculation that leads to
\eqref{modular}. To do so, we  
will exploit the fact that the partition function 
\beq
Z_0(\xi) = \prod_{j=1}^\infty \left(1-e^{-\xi j}\right)^{-1} = (1-e^{-\xi})^{-1}(1-e^{-2\xi})^{-1}
(1-e^{-3\xi})^{-1}\ldots
\eeq
has the following remarkable property, which is obtained using the
Poisson summation formula: 
\beq
Z_0(\xi) = \sqrt{\frac\xi{2\pi}}e^{\frac{\pi^2}{6\xi} - \frac{\xi}{24}}Z_0(4\pi^2/\xi).
\label{prop}
\eeq
The first step is to write the partition function we are interested in ($\xi=\beta\hbar/2l$),
\beq
Z(\xi) = e^{-2lM_0\xi/\hbar} \prod_{j=0}^\infty\left[1-e^{-\xi(2j+1)}\right]^{-1} = 
e^{-2lM_0\xi/\hbar} (1-e^{-\xi})^{-1}(1-e^{-3\xi})^{-1}(1-e^{-5\xi})^{-1}\ldots,
\eeq
in terms of $Z_0(\xi)$. In fact, it is clear that
\beq
Z(\xi) = e^{-2lM_0\xi/\hbar} \frac{Z_0(\xi)}{Z_0(2\xi)}.
\eeq
Now using the property \eqref{prop} of $Z_0(\xi)$, we find
\bea
K(\xi) &\equiv& \frac{Z_0(\xi)}{Z_0(2\xi)} = \sqrt{\frac\xi{2\pi}} e^{\frac{\pi^2}{6\xi}-\frac{\xi}{24}}
\times \sqrt{\frac\pi\xi}e^{-\frac{\pi^2}{12\xi}+\frac\xi{12}}\frac{Z_0(4\pi^2/\xi)}
{Z_0(2\pi^2/\xi)}\cr\cr
&=& \frac 1{\sqrt{2}}~e^{\frac{\pi^2}{12\xi}+\frac{\xi}{24}}
[K(2\pi^2/\xi)]^{-1}, 
\eea
implying that
\beq
Z(\xi) = \frac 1{\sqrt{2}}~e^{-2lM_0\xi/\hbar +
  \frac{\pi^2}{12\xi}+\frac{\xi}{24}-\frac{4\pi^2lM_0}{\xi\hbar}} 
[Z(2\pi^2/\xi)]^{-1}, 
\eeq
which is \eqref{modular}.
\bigskip\bigskip

\centerline{\bf APPENDIX B}
\bigskip

If the shells obey Boltzmann statistics, then 
\beq
g(\cN_1,\ldots,\cN_j,\ldots) = \frac{\cN!}{\cN_1!\ldots\cN_j!\ldots}
\eeq
gives the number of distinguishable rearrangements of a particular distribution 
of $\cN$ shells between the energy levels. In this case,
\beq
Z(\beta) = \sum_{\{\cN_1,\ldots,\cN_j,\ldots\}}\frac{\cN!}{\cN_1!\ldots
\cN_j!\ldots} \exp\left[-\beta \left(M_0+\sum_{j}\mu_j \cN_j\right)\right],
\eeq
and the sum is to be evaluated subject to the requirement that the total number 
of shells is fixed to be $\cN$. Thus we find
\beq
Z(\beta) = 2^{-\cN} e^{-\beta M_0} \sinh^{-\cN} \left(\frac{\beta\hbar}{2l}\right).
\eeq
The average energy in the limit as $\beta \rightarrow 0$ is easily found to be 
$M \approx M_0+\cN/\beta$. However, the dependence of the inverse
temperature on  
the black hole mass is considerably different from \eqref{entemp}; we
now find the entropy to be
\beq
S \approx \cN + \cN \ln \left[\frac{(M-M_0)l}{\cN\hbar}\right].
\label{ent2}
\eeq
Its maximum value of $S\approx (M-M_0)l/\hbar$ is achieved when the number of shells $\cN 
\approx (M-M_0)l/\hbar$.
This answer can also be obtained directly in the microcanonical ensemble by counting the 
number of ways in which $Q$ quanta can be distributed among $\cN$ shells, 
\beq
\Omega(Q,\cN) = \frac{(\cN+Q-1)!}{(\cN-1)!Q!}.
\label{microcount}
\eeq
In the high temperature limit, and keeping in mind that $(M-M_0)l/\hbar = Q + \cN/2$, we obtain 
precisely \eqref{ent2}. It does not agree with the Bekenstein--Hawking entropy in its 
dependence on the mass of the black hole.

\end{document}